# Bistochastically private release of longitudinal data

Nicolas Ruiz

Universitat Rovira i Virgili
Departament d'Enginyeria Informàtica i Matemàtiques
Av. Països Catalans 26, 43007 Tarragona, Catalonia
`nicolas.ruiz@urv.cat`

**Abstract.** Although the bulk of the research in privacy and statistical disclosure control is designed for cross-sectional data, i.e. data where individuals are observed at one single point in time, longitudinal data, i.e. individuals observed over multiple periods, are increasingly collected. Such data enhance undoubtedly the possibility of statistical analysis compared to cross-sectional data, but also come with one additional layer of information, individual trajectories, that must remain practically useful in a privacy-preserving way. Few extensions, essentially *k*-anonymity based, of popular privacy tools have been proposed to deal with the challenges posed by longitudinal data, and these proposals are often complex. By considering randomized response, and specifically its recent bistochastic extension, in the context of longitudinal data, this paper proposes a simple approach for their anonymization. After having characterized new results on bistochastic matrices, we show that a simple relationship exists between the protection of each data set released at each period, and the protection of individuals' trajectories over time. In turn, this relationship can be tuned according to desired protection and information requirements. We illustrate the application of the proposed approach by an empirical example.

**Keywords.** Privacy, anonymization, longitudinal data, randomized response, bistochastic matrices

## 1    Introduction

To manage the tension between data collection and exploratory data analytics on the one hand, and stronger data protection legislation on the other hand, *anonymization* stands out to reconcile both sides, by providing effective privacy protection while offering suitable data utility. Most anonymization techniques and privacy models available today are designed for static data, observed all at once, not revised and published only one time. However, there are several types of individual data that can be published in a privacy-preserving way for fulfilling analysis needs, e.g. relational data, transaction data, stream data, trajectory data … These data types differ in structure, properties and the information they contain about individuals. The dissemination of any specific type entails its own privacy risks and information preservation requirements, which should ideally be considered by the anonymization approach selected to protect the data.

Among these different types, longitudinal data are of particular interest in many areas, e.g. economics, medical research, sociology, finance, marketing... Longitudinal data are repeated observations of generally the same respondents that are published at different points in time. They vary from cross-sectional data, i.e. where individuals are observed at a single point in time, from time-series data, i.e. where *one* single entity is observed along generally a long-time span, and from stream data, i.e. where new individuals are continuously added to the data set. The defining feature of longitudinal data is that the multiple observations within the same individuals can be ordered across time. Longitudinal surveys generally use calendar time, months or years, as the dimension separating observations on the same subject. Although the notion of time in longitudinal data can be quite intricate [1], in this paper we will focus on repeatedly measured attributes over the same individuals that can be ordered along a line to describe the sequence of measurement.

**Contribution and plan of this paper**

Even though the anonymization literature offers a wide variety of tools suited to different contexts and data types [2], there have been relatively few attempts to deal with the challenges posed by longitudinal data. Current proposals, formulated generally in the context of medical data, are based on global suppression, generalization, and/or clustering, to deliver *k*-anonymity-related privacy guarantees (e.g. [3], [4], [5]). Moreover, these proposals almost exclusively focus on the protection of individuals' trajectories by converting, prior to performing anonymization, longitudinal data into multi-dimensional sequential data sets (see [5] for an outline of such procedure). This step, combined with the inherent complexities of the anonymization methods proposed, make the whole task of protecting longitudinal data intricate.

Moreover, those proposals remain mute on the fact that longitudinal data are repeated cross-sectional data which happen to follow the same set of individuals. That is, each data published at each period can be viewed, in isolation, as cross-sectional data requiring their own protections, and then additional protection needs to be implemented on the longitudinal information, i.e. the individuals' trajectories. For instance, in practice this is the case for the EU Statistics on Income and Living Conditions [6], where individuals surveyed at time $t$ are released as a cross-sectional data set, and then a subset of these individuals are randomly selected in $t+1$ to be surveyed again. In that case, by focusing only on trajectories, one is discarding protection at the cross-sectional level.

Finally, and as data releases are accumulating over time, the privacy properties will most likely need to be re-evaluated, a cumbersome task. To alleviate this last constraint, current proposals assume that all released data have been observed before proceeding to anonymization, which could appear as unrealistic.

Based on this state of affairs, the purpose of this paper is to break ground towards a simple approach for the anonymization of longitudinal data that operates both at the cross-sectional and longitudinal levels, by establishing the links between the two, in particular how anonymization applied to the former translates to the latter. To achieve this, we consider randomized response, a well-established mechanism with rigorous privacy guarantees [7]. To the best of our knowledge, this is the first reported work that considers randomized response in the context of longitudinal data.

Using the recently proposed bistochastic version of randomized response [8], which connects several fields of the privacy literature, and after having characterized new theoretical results on bistochastic matrices, we show that, when each cross-sectional data set at time *t* is protected, as in the classic static data scenario, it exists a simple relationship between this protection and the protection of individuals' trajectories over time. In turn, this relationship can be tuned according to desired protection and information requirements. Moreover, not all data releases need to be observed to conduct anonymization.

The remainder of this paper is organized as follows. Section 2 gives some background elements on randomized response and bistochastic privacy, needed later. Section 3 develops the main approach proposed in this paper, underpinned by new theoretical insights. Section 4 presents some empirical results based on this new approach. Conclusions and future research directions are gathered in Section 5.

## 2   Background elements

### 2.1   Randomized response

Let *X* denotes an original categorical attribute with $1, \dots, r$ categories, and *Y* its anonymized version. Given a value $X = u$, randomized response (RR, [7]) computes a value $Y = v$ by using an $r \times r$ Markov transition matrix:

$$P = \begin{pmatrix} p_{11} & \cdots & p_{1r} \\ \vdots & \ddots & \vdots \\ p_{r1} & \cdots & p_{rr} \end{pmatrix} \quad (1)$$

where $p_{uv} = \Pr(Y = v | X = u)$ denotes the probability that the original response *u* in *X* is reported as *v* in *Y*, for $u, v \in \{1, \dots, r\}$. To be a proper Markov transition matrix, it must hold that $\sum_{v=1}^{r} p_{uv} = 1 \; \forall u = 1, \dots, r$. *P* is thus *right stochastic*, meaning that any original category must be spread along the anonymized categories.

The usual setting in RR is that each subject computes her randomized response *Y* to be reported instead of her true response *X*. This is called the *ex-ante* or local anonymization mode. Nevertheless, it is also possible for a (trusted) data curator to gather the original responses from the subjects and randomize them in a centralized way. This *ex-post* mode corresponds to the Post-Randomization Method (PRAM, [9]). Apart from who performs the anonymization, RR and PRAM operate the same way and make use of the same matrix *P*.

RR is based on an implicit privacy guarantee called *plausible deniability* [10]. It equips the individuals with the ability to deny, with variable strength according to the parameterization of *P,* that they have reported a specific value. In fact, the more similar the probabilities in *P*, the higher the deniability. In the case where the probabilities within each column of *P* are identical, it can be proved that *perfect secrecy* in the Shannon sense is reached [11]: observing the anonymized attribute *Y* gives no information at all on the real value *X*. Under such parameterization of *P*, a privacy breach cannot originate from the release of an anonymized data set, as the release does not bring any information that could be used for an attack. However, the price to pay in terms of data utility is high [10].

## 2.2  Bistochastic privacy

We will assume now that the randomized response matrix P above fulfills the additional left stochasticity constraints that $\sum_{u=1}^{r} p_{uv} = 1 \ \forall v = 1, \ldots, r$. This makes P *bistochastic* (left stochasticity implying that any anonymized categories must come from the original categories).

At first sight, one could wonder about the necessity of imposing an additional constraint on RR and its ex-post version PRAM, some well-established approaches for anonymization that have proved their merits over the years. However, it happens that the bistochasticity assumption connects several fields of the privacy literature, including the two most popular models, *k*-anonymity and ε-differential privacy, but also any Statistical Disclosure Control (SDC) method. Indeed, it can be shown that *k*-anonymity guarantees can be reached using a block-diagonal, bistochastic matrix where each block achieves perfect privacy, while ε-differential privacy guarantees can be reached using a circulant, bistochastic matrix. Additionally, it can also be shown that any SDC method can be viewed as a specific case of a more general approach that uses bistochastic matrices to perform anonymization. We refer the reader to [10] for a detailed presentation of these results.

Beyond its unifying properties, the bistochastic version of randomized response offers additional advantages by clarifying and operationalizing the trade-off between protection and utility. Indeed, it is well-known that a bistochastic matrix never decreases uncertainty and is the only class of matrices to do so [12]. Stated otherwise, when a bistochastic P is applied to an original attribute X, its anonymized version Y will always contain more entropy than X. Remark that when P is only right stochastic, as in the traditional approach to RR, no particular relationship emerges.

Now, and as a direct consequence of this last property, the strength of anonymization (and equivalently the strength of plausible deniability), can be measured, as in cryptography for the strength of security, in terms of bits, through $H(P)$, the entropy rate of P (this rate being the average of the entropies of each row of P [13]). The entropy rate is a central quantity in bistochastic privacy, as it measures the formal privacy guarantee (and the information content) that equipes a bistochastically private data set.

In the case of perfect secrecy where all probabilities in P are equal, and that we will denote hereafter by $P^*$, we have $H(P^*) = \log_2 r$, which is the maximum achievable entropy for an $r \times r$ bistochastic matrix. Thus, after the choice of a suitable parameterization, the number of bits that P contains establishes a metric in terms of plausible deniability.

From these results, the definition of bistochastic privacy follows. We provide here the univariate case, where we seek to anonymize only one attribute to prevent disclosure (other definitions at the data set level can be found in [10]):

***Definition 1 (Univariate bistochastic privacy)***: *The anonymized version Y of an original attribute X is β-bistochastically private for $0 \leq \beta \leq 1$ if and only if:*
  i)   $Y = P'X$ with P bistochastic
  ii)   $\frac{H(P)}{H(P^*)} \geq \beta$.

### 2.3 Static vs. longitudinal data

Compared to cross-sectional data, longitudinal data provide some clear advantages as they are generally more informative, by capturing individuals' trajectories along variables of interest. Indeed, cross-sectional distributions that look relatively stable can in fact hide a multitude of changes that can only be captured if the same set of individuals is followed over time. For example, spells of unemployment, job turnover, residential and income mobility are better studied with longitudinal data. Longitudinal data are also well suited to study states durations, e.g. disease, unemployment and poverty, and if the time dimension is long enough, they can shed light on the speed of adjustments to medical treatments or policy changes. For instance, in measuring unemployment, cross-sectional data can estimate what proportion of the population is unemployed at a point in time. Repeated cross-sections can show how this proportion changes over time. But only longitudinal data can estimate what proportion of those who are unemployed in one period can remain unemployed in another period.

Now, it is clear that longitudinal data comes with specific privacy challenges. While it is beyond the scope of the present paper to investigate exhaustively the possible forms of an attacker's background knowledge specific to longitudinal data, we can outline the main ones. Indeed, such knowledge may be thought of with its own characteristics compared to other types of data, and in particular cross-sectional data. For example, an adversary may know that someone has transitioned from unemployment to employment between two time periods. Thus, the employment status trajectory adds information to a cross-sectional value of the employment status, and hence increases the re-identification power of employment status when considered as a quasi-identifier.

Along the same line, changes in confidential attributes, such as salary, can also be viewed as a quasi-identifier: an attacker may for example not know the salary of an individual at two periods, but may know that it has increased significantly between the two and can use that information to conduct the attack. Thus, the individual may consider as a privacy risk the fact that someone can learn about his salary trajectory, even if his salaries at the two time periods are not disclosed, e.g. the two salary values have been masked enough to avoid attribute disclosure, but the masked values did not alter the salary trajectory, providing insights for the intruder. Thus, it can be reasonably stated that longitudinal data generally widen privacy threats. However, this widening is also a widening of information specific to longitudinal data. This is in fact what make them specifically valuable in the first place and must be preserved to a lesser or greater extent for the dissemination of longitudinal data to be useful.

## 3 Bistochastic randomization of longitudinal data

In what follows, we will first consider the case of one single categorical attribute with $r$ categories, observed over the same set of individuals over time. Abusing slightly the usual terminology, for convenience we will call each release at time $t$ a cross-section, and the collection of cross-sections released up to time $T$ a longitudinal data set.

We will also assume that the data are collected over periods $t=1,...,T$ with $T$ finite but as large as necessary.

### 3.1 Illustrative example of the proposed approach

Let us first consider a basic example with $r=2$ and $T=2$. This binary attribute is anonymized in $t=1$ with the bistochastic matrix $\begin{pmatrix} 0.9 & 0.1 \\ 0.1 & 0.9 \end{pmatrix}$ and in $t=2$ with $\begin{pmatrix} 0.7 & 0.3 \\ 0.3 & 0.7 \end{pmatrix}$. If we abstract momentarily from the longitudinal nature of the data and seek to measure the overall level of protection for the attribute over the two periods, one convenient way to represent this is through the following block diagonal matrix:

$$\begin{pmatrix} 0.9 & 0.1 & 0 & 0 \\ 0.1 & 0.9 & 0 & 0 \\ 0 & 0 & 0.7 & 0.3 \\ 0 & 0 & 0.3 & 0.7 \end{pmatrix}$$

The entropy rate of this matrix will provide the overall level of protection applied to the attribute's values in $t=1$ and then in $t=2$ as if they were not about the same individuals, that is as if they were two static data sets. Now, as discussed above the central feature of longitudinal data is that attributes are followed over time for the same set of individuals, and individuals' trajectories must also be protected. We remark that the protection applied to these trajectories can be represented by the Kronecker product of the two bistochastic matrices:

$$\begin{pmatrix} 0.9 & 0.1 \\ 0.1 & 0.9 \end{pmatrix} \otimes \begin{pmatrix} 0.7 & 0.3 \\ 0.3 & 0.7 \end{pmatrix} = \begin{pmatrix} 0.63 & 0.27 & 0.07 & 0.03 \\ 0.27 & 0.63 & 0.03 & 0.07 \\ 0.07 & 0.03 & 0.63 & 0.27 \\ 0.03 & 0.7 & 0.27 & 0.63 \end{pmatrix}$$

For instance, the first term of the matrix means that an individual having declared to be in the first category in both $t=1$ and $t=2$ has a 0.63 probability of being reported as having the same trajectory in the protected longitudinal data set. Similarly, if that individual reported to be in the first category in t=1 and in the second in $t=2$, as indicated by the second diagonal term, it has also a 0.63 probability of being reported as having the same trajectory after protection. Now, considering the second probability in the first column, an individual being in the first category in both $t=1$ and $t=2$ has a 0.27 probability to be reported as being in the first category in $t=1$, but in the second in $t=2$; her trajectory has been modified by anonymization. Moreover, recalling the following Theorem, this Kronecker product is also bistochastic:

***Theorem 1 (Marshall, Olkin, and Arnold [12]):** If P and Q are m×m and n×n bistochastic matrices, respectively, then the Kronecker product $P \otimes Q$ is an mn × mn bistochastic matrix.*

Thus, the Kronecker product can be interpreted in the same way as any stochastic matrices used for RR. For instance, its diagonal contains the probability that the anonymized trajectories are the true ones, meaning that the diagonal values will indicate a certain level of "truthfulness" in the trajectories, similarly to RR applied to a cross-sectional attribute, see *Equation (1)*.

Now, by the associative property of the Kronecker product, and using repeatedly *Theorem 1*, the Kronecker product of an arbitrarily finite number of bistochastic matrices will also be bistochastic. We can thus repeat this procedure as *T* grows. However, the resulting product will quickly grow out of control to be reasonably tractable, not least for an attribute with a large number of categories. In what follows, we establish new results linking the entropy rate of block diagonal matrices with the entropy rate of Kronecker products of bistochastic matrices. Those results will then allow us to establish a simple relationship for the anonymization of longitudinal data.

### 3.2 Results on the entropy rates of block diagonal matrices and Kronecker products using bistochastic matrices as components

***Theorem 2:*** *If $P_1,...,P_T$ are $n_1 \times n_1, ..., n_T \times n_T$ bistochastic matrices with entropy rates $H(P_1),...,H(P_T)$, respectively, then the entropy rate of the block diagonal matrix* $\begin{pmatrix} P_1 & & 0 \\ & \ddots & \\ 0 & & P_T \end{pmatrix}$ *is the weighted average of the entropy of each sub-block, with weights equal to the relative size of the sub-blocks.*

***Proof:*** *Assume P a $N \times N$ block diagonal matrix with T bistochastic blocks $P_1,...,P_T$ of size $n_1 \times n_1,..., n_T \times n_T$, respectively, and $N = \sum_{t=1}^{T} n_t$. Since P is by construction also bistochastic, its stationary $\pi$ distribution is the uniform one, with $\pi_i = \frac{1}{N} \forall i = \{1,...,N\}$. The entropy rate of P is thus:*

$$H(P) = -\sum_{i,j} \pi_i p_{i,j} \log_2 p_{i,j} = -\frac{1}{N} \sum_{i,j} p_{i,j} \log_2 p_{i,j}.$$

*Since P is block diagonal, transitions in this Markov chain occur only within blocks. We can thus partition the entropy rate over states i,j into contributions from each block $P_t$:*

$$H(P) = -\frac{1}{N} \sum_{t=1}^{T} \sum_{i,j \in P_t} p_{i,j} \log_2 p_{i,j}$$

*For each block $P_t$, the entropy rate is defined as:*

$$H(P_t) = -\frac{1}{n_t} \sum_{i,j \in P_t} p_{i,j} \log_2 p_{i,j}$$

*which implies:*

$$\sum_{i,j \in P_t} p_{i,j} \log_2 p_{i,j} = -n_t H(P_t)$$

*Substituting this expression into $H(P)$, one gets:*

$$H(P) = \sum_{t=1}^{T} \frac{n_t}{N} H(P_t) \quad (2)$$

*The overall entropy rate of the block diagonal bistochastic matrix is the weighted average of the entropy rates of its blocks, where the weights are determined by the relative sizes of the blocks $\frac{n_t}{N}$.*

**Theorem 3:** If $P_1, ..., P_T$ are $n_1 \times n_1, ..., n_T \times n_T$ bistochastic matrices with entropy rates $H(P_1), ..., H(P_T)$, respectively, then the entropy rate of the Kronecker product $P_1 \otimes ... \otimes P_T$ is the sum of the entropy rates of each matrix composing the product.

*Proof:* Consider the first two terms in $P_1 \otimes ... \otimes P_T$. $P_1$ and $P_2$ have $n_1$ and $n_2$ states, respectively, with entropy rates $H(P_1) = -\sum_{i,k} n_1 p_{i,k} \log_2 p_{i,k}$ and $H(P_2) = -\sum_{i,k} n_2 p_{j,l} \log_2 p_{j,l}$, respectively. Denoting the transition probabilities of $P_1 \otimes P_2$ by $(P_1 \otimes P_2)((i,j),(k,l))$, we have $(P_1 \otimes P_2)((i,j),(k,l)) = p_{i,k} p_{j,l}$.

By Theorem 1, the result of $P_1 \otimes P_2$ is bistochastic, with as stationary distribution the uniform one. We can thus compute its entropy rate, which is given by:

$$H(P_1 \otimes P_2) = -\frac{1}{n_1 n_2} \sum_{i,j,k,l} p_{i,k} p_{j,l} \log_2 p_{i,k} p_{j,l}$$

*Developing this term above using logarithmic properties, we get:*

$$H(P_1 \otimes P_2) = -\frac{1}{n_1 n_2} \left[ \sum_{i,k} p_{i,k} \log_2 p_{i,k} \sum_{j,l} p_{j,l} + \sum_{j,l} p_{j,l} \log_2 p_{j,l} \sum_{i,k} p_{i,k} \right]$$

Since $P_1$ and $P_2$ are bistochastic, it holds that $\sum_{j,l} p_{j,l} = n_2$ and $\sum_{i,k} p_{i,k} = n_1$. Simplifying, the first term in the bracket becomes $-n_1 n_2 H(P_1)$, while the second becomes $-n_2 n_1 H(P_2)$. Reinjecting those terms in the above expression, one gets:

$$H(P_1 \otimes P_2) = -\frac{1}{n_1 n_2}[-n_1 n_2 H(P_1) + -n_2 n_1 H(P_2)] = H(P_1) + H(P_2)$$

Consider now $P_1 \otimes P_2 \otimes P_3$. By remarking again that the resulting matrix will be bistochastic by Theorem 1, and repeating the same as above, this necessarily leads to $H(P_1 \otimes P_2 \otimes P_3) = H(P_1) + H(P_2) + H(P_3)$. Repeating up to T, we have:

$$H(P_1 \otimes ... \otimes P_T) = \sum_{t=1}^{T} H(P_t) \quad (3)$$

*Theorem 3* leads to a remarkably simple results which, to the best of our knowledge, has never been characterized before in the majorization literature (as well as Theorem 2; see [XX] for a comprehensive survey in this field). However, this result follows intuitively from the fact that the Kronecker product models independent transition across sub-markov chains, and entropy is always additive under independence.

### 3.3 Implications for longitudinal data anonymization

Assuming now that the attribute's categorization does not change over time (see more on this assumption below), i.e. $n_1=...=n_T=n$ *Equation (2)* becomes $H(P) = \frac{1}{T}\sum_{t=1}^{T} H(P_t)$. Combining with *Equation (3)*, we have:

$$H(P_1 \otimes ... \otimes P_T) = T.H(P) \quad (4)$$

*Equation (4)* states that the protection of trajectories in a longitudinal data set (as conveyed by the entropy rate in bistochastic privacy) is the number of time periods in the overall longitudinal data set multiplied by the average entropy rate applied *as if each release is considered as purely cross-sectional*. This result is possible by use of the bistochasticity assumption. Beyond the various consequences mentioned above arising from this assumption for static data sets, for longitudinal data it also connects in a simple way cross-sectional and trajectory anonymization.

Several consequences from this result are worth noting. First, a simple and rather expected relationship emerges for the protection of longitudinal data: *the more (less) protection is applied to cross-sections, the more (less) protection is applied to trajectories*. Moreover, and because of the connective properties of bistochastic privacy with *k*-anonymity and differential privacy, the same applies to these privacy models. Such relationship has never been characterized in past contributions due to the approaches selected, which were focusing mainly on trajectories.

Second, we believe the approach adopted in this paper, RR with the bistochasticity assumption, considerably simplifies the practice of longitudinal data anonymization. In fact, the privacy practitioner can simply protect each cross-section as they come, and from this task monitor with *Equation (3)* the protection level of trajectories, subsequently adjusting it by more or less protecting coming cross-sections (or revising the protection of past ones). Through this approach, *the task of anonymizing a longitudinal data set is equivalent to anonymizing static data sets, except for one additional quantity to scrutinize*. Doing so, the entailed complexity boils down to the standard and tractable complexity of randomization, being conducted *ex-ante* or *ex-post*. This departs from current proposals in the literature, which require a pre-treatment of the longitudinal data set, plus an additional complexity inherent to the methodologies proposed. In particular, *Equation (3)* allows grasping with the protection of trajectories without having to create a multi-dimensional sequential data set (nor do we need to compute the Kronecker product of the cross-sectional randomization matrices, as what needs only to be observed is its resulting entropy rate).

Third, this approach does not require to observe all time periods to perform protection, as is necessary in other proposed procedures. Instead, one can adjust the protection of trajectories as new data are coming in.

Finally, beyond *Equation (4)*, and bearing in mind that in bistochastic privacy entropy is related to the maximum injectable entropy, we can define longitudinal bistochastic privacy guarantees for an attribute, where at each period each release is bistochastically private, and also the trajectories (thus starting in *t*=2 for such trajectories to exist):

***Definition 2 (Univariate longitudinal bistochastic privacy)***: *For T≥2, the anonymized version Y=(Y$_2$,…, Y$_T$) of an original attribute released over T periods X=(X$_2$,…, X$_T$) is $\beta_{T,L}$-bistochastically private with $\beta_{T,L} = ((\beta_2, …, \beta_T), \beta_L)$ for $0 \leq \beta_t \leq 1 \, \forall t = 2, …, T$ and $0 \leq \beta_L \leq 1$, if and only if :*
  i)   $Y_t = P_t' X_t$ with $P_t$ bistochastic $\forall t = 2, …, T$
  ii)  $\frac{H(P_t)}{H(P_t^*)} \geq \beta_t \, \forall t = 2, …, T.$
  iii) $\frac{\sum_{t=2}^{T} H(P_t)}{\sum_{t=2}^{T} H(P_t^*)} \geq \beta_L$

### 3.4 Extensions

**Numerical attributes and balanced *vs*. non-balanced longitudinal data**

Nothing precludes, conceptually or practically, to apply bistochastic matrices on a numerical attribute ([10], [14]). In the categorical case, the original proportions of respondents whose values fall in each of the $r$ categories will be changed, which will coarsen the distribution to deliver randomized proportions closer to the uniform distribution. In the numerical case, it will tend to average the numerical values of respondents. In fact, if randomization is performed *ex-post*, i.e. using PRAM [9], then in a bistochastic randomized response scheme on a numerical attribute *the individuals are used as categories*. Moreover, and because bistochastic matrices are mean-preserving, the anonymized numerical attribute will have the same mean as the original numerical attribute (note that this would not be possible with a non-bistochastic Markov matrix, which is generally not mean-preserving).

However, the application on numerical attributes raises the issue of balanced *vs*. non-balanced longitudinal data. It is well known that longitudinal data can be subject to varying rates of nonresponse from subjects over time, an issue generally leading to what is called an unbalanced longitudinal data set [15], i.e. *not every* subject is observed at *every* period. In the categorical case, as far as performing privacy is concerned this is not problematic, as *randomization is happening across a pre-determined categorization*, which is unlikely to change over time. In fact, if it were to change, longitudinal data would lose their appeal by being considerably less informative on trajectories.

However, for a numerical attribute it will, by nature, be modified. For privacy this becomes a concern as the size of the numerical attribute will vary over time. The results developed above can incorporate this, especially as *Theorems 2 and 3* have been developed with bistochastic matrices of different sizes. However, *Equation 4* will not hold anymore. This can be easily recovered by setting the size of the bistochastic matrices as observed in $t=1$ and then, when some subjects are not replying in latter periods, transition probabilities for those individuals are simply put to zero. Remark here that we are not considering the possibility of new subjects being added, which is a central feature of data stream but not of longitudinal data [14].

**The case of several attributes**

We now consider a longitudinal data set with $M$ attributes $(A_{1,t}, \ldots, A_{M,t})$., being numerical and/or categorical, all collected over periods $t=1,\ldots,T$. In that context, following [10], at each period bistochastic randomized response can be applied on the joint distribution $A_{1,t} \times \ldots \times A_{M,t}$. Indeed, it is apparent that, conceptually, nothing precludes to strictly apply the procedure proposed in this paper. Practically however, this is not without limitations. To perform anonymization at a given period, the implied bistochastic matrix may reach a very large size, in particular if the original data set contains many numerical attributes. Remark that in that case, the Kronecker product of these large matrices becomes unreasonably large but, once again, this product does not need to be directly computed. The practical hurdle here resides more in the randomization at each period and, like other privacy models, bistochastic privacy is not immune to the curse of dimensionality [16].

## 4 Empirical illustrations

We start by noting that, to achieve bistochastic privacy, one just needs to parameterize bistochastic matrices, as one does not need to observe the actual data [10]. Therefore, an agent, independent of the data controller, say a "data protector", can generate the appropriate matrices. The associated parameter $\beta$ for those matrices will depend on the environment and the desired protection-utility trade-off. As a result, performing bistochastic privacy can be viewed as close to performing cryptography, where an encryption key is generated independently of the message to be protected.

In the example below, we act as a data protector who generates bistochastic matrices to protect $r=100$ individuals, each with one value for a numerical attribute observed during $T=5$ periods. Following [17], the data protector will parameterize bistochastic matrices to reach differential privacy guarantees for various $\varepsilon$ values. She will protect each release as if they are purely cross-sectional and will also check the protection's level of the trajectories (i.e. the entropy level of the Kronecker product of the bistochastic matrices used at each period). The results of this exercise are shown in Table 1.

**Table 1.** Example of bistochastic privacy guarantees on longitudinal data.

| Epsilon values at each period | Bistochastically private guarantees (for each period and trajectories) | | | | | | | | |
|---|---|---|---|---|---|---|---|---|---|
| | T=1 | T=2 | | T=3 | | T=4 | | T=5 | |
| | Cross section | Cross section | Trajectories | Cross section | Trajectories | Cross section | Trajectories | Cross section | Trajectories |
| {2;2;2;2;2} | 70% | 70% | 70% | 70% | 70% | 70% | 70% | 70% | 70% |
| {0.1;0.1;0.1;0.1;0.1;0.1} | 93% | 93% | 93% | 93% | 93% | 93% | 93% | 93% | 93% |
| {3;2;1.5;1;0.5} | 42% | 70% | 56% | 83% | 64% | 93% | 72% | 98% | 78% |
| {1;0.8;0.5;0.3;0.1} | 93% | 95% | 93% | 98% | 95% | 99% | 96% | 99% | 97% |

In this example, under the first two sets of epsilon values, where the data protector decides not to modify the level of protection across time, the trajectories' protection is equal to the level for each cross-section, which follows from *Definition 2*. Also expected is the fact that the second set is offering more protection, because of the lower epsilon values. For the other two sets, where the data protector modifies the level of protection over time, toward more privacy, the protection's levels of the trajectories also increase over time.

Remark that this example leverages the connective properties of bistochastic privacy, using differential privacy to parameterize the bistochastic matrices. What matters in bistochastic privacy is how much entropy is injected (and in fact, this amount can be set independently of already existing privacy models), as a percentage of the maximum injectable entropy (i.e. the perfect privacy case, see *Definition 1*). Here, it happens also that each cross-section is also equipped with interpretable differential privacy guarantees. However, this does not translate to the protection of trajectories. In fact, it can be shown that this is not the case, as the Kronecker product of bistochastic matrices with differential privacy guarantees does not exhibit itself those guarantees. As such, the protection of trajectories displayed in Table 1 should be interpreted only from the bistochastic privacy lens.

Similarly, for each cross-section we could have parameterized bistochastic matrices using *k*-anonymity guarantees [10]. However, in that case it can be shown that the trajectories also exhibit *k*-anonymity guarantees. We do not report those results here due to space constraints, but they will be made available in an extended version of this paper.

## 5      Conclusions and future work

This paper considered bistochastic privacy, a recently proposed type of randomized response, in the context of longitudinal data anonymization. After having characterized new theoretical results on bistochastic matrices, we established a simple link between anonymization at the cross-sectional and longitudinal levels, making the anonymization of longitudinal data sets simple because equivalent in practice to the anonymization of static data sets, except for one additional quantity to consider. Each release of a longitudinal data at given period set can be protected with rigorous privacy guarantees as if it was static in nature, and then the protection of trajectories, also with rigorous privacy guarantees, can be monitored and eventually adjusted through coming new releases or revisions of past ones.

This paper opens several lines for future research. One of them is to introduce the notion of privacy budget in the present framework, to devise binding rules in terms of entropy for coming new releases when an overall level of protection/information for the longitudinal data set is targeted. Conducting further empirical work on real-life longitudinal data is also warranted. Another path is to consider republishing strategies for longitudinal data, which entail additional privacy risks [18]. Yet another challenge is to investigate the compatibility, within the framework proposed in this paper, of the recent solutions to mitigate the dimensionality problem in randomized response [16].